\documentclass{article}

\usepackage{amsmath,amsfonts,bm}

\def\eqref#1{equation~(\ref{#1})}

\def\1{\bm{1}}

\DeclareMathAlphabet{\mathsfit}{\encodingdefault}{\sfdefault}{m}{sl}
\SetMathAlphabet{\mathsfit}{bold}{\encodingdefault}{\sfdefault}{bx}{n}

\usepackage{spconf,amsmath,graphicx}
\usepackage{algpseudocode}
\usepackage[ruled,vlined,linesnumbered]{algorithm2e}    
\usepackage{wrapfig}
\usepackage{amssymb}
\usepackage{hyperref}
\usepackage{caption}
\usepackage{pifont}

\newcommand{\myparagraph}[1]{\noindent\textbf{#1}\hspace{.25cm}}

\title{A Vector Quantized Masked AutoEncoder for speech emotion recognition}
\name{Samir Sadok, Simon Leglaive, Renaud Séguier}
\address{CentraleSup\'elec, IETR UMR CNRS 6164, France}
\begin{document}
\ninept
\maketitle
\begin{abstract}
Recent years have seen remarkable progress in speech emotion recognition (SER), thanks to advances in deep learning techniques. However, the limited availability of labeled data remains a significant challenge in the field. Self-supervised learning has recently emerged as a promising solution to address this challenge. In this paper, we propose the vector quantized masked autoencoder for speech (VQ-MAE-S), a self-supervised model that is fine-tuned to recognize emotions from speech signals. The VQ-MAE-S model is based on a masked autoencoder (MAE) that operates in the discrete latent space of a vector quantized variational autoencoder. Experimental results show that the proposed VQ-MAE-S model, pre-trained on the VoxCeleb2 dataset and fine-tuned on emotional speech data, outperforms an MAE working on the  raw spectrogram representation and other state-of-the-art methods in SER.
\end{abstract}
%
\section{Introduction}
\label{sec:intro}
Speech emotion recognition (SER) \cite{el2011survey} is a research area focused on automatically identifying emotions from speech signals. With the growth of technology and the increasing use of speech-based interfaces, there is a growing demand for systems that can accurately recognize emotions in speech. In recent years, deep learning methods have played a major role in improving SER performance \cite{akccay2020speech}. However, the scarcity and high cost of obtaining labeled speech data for emotion recognition pose a major difficulty. To address this challenge, researchers have shifted their focus towards self-supervised learning approaches \cite{wang2021fine, chen2021exploring}. In these approaches, models are pre-trained on a self-supervised task, such as predicting masked tokens in speech signals, and then fine-tuned on a smaller set of labeled data for the SER task \cite{gong2022ssast, pepino2021emotion}. This method has the advantage of scalability, as the self-supervised task can be trained on large amounts of unlabeled speech data, reducing the need for labeled data \cite{liu2022audio, zhang2022survey}. Self-supervised training has been successfully applied in the field of SER, and has demonstrated promising results by allowing the model to learn useful representations of speech signals for emotion recognition, even in scenarios where labeled data is limited \cite{pepino2021emotion, macary2021use}.

This article focuses on self-supervised SER with the masked autoencoder (MAE) approach \cite{he2022masked}. The MAE is an asymmetrical encoder-decoder architecture that relies on input masking \cite{he2022masked}. Originally developed in natural language processing (NLP) \cite{devlin2018bert}, the MAE approach has also been applied to image analysis using vision transformers (ViT) \cite{dosovitskiy2020image}. The MAE process involves dividing the input into non-overlapping patches, each represented by a token embedding. A large proportion of tokens are masked (usually 75\% for image modeling and 15\% for text modeling), and only the visible tokens are fed to the encoder. A lightweight decoder then reconstructs the image/text by combining the visible tokens from the encoder and learnable mask tokens. The cost function is applied only to the masked tokens. Recently, the MAE has been adapted for audio using 2D time-frequency representations such as the mel-spectrogram \cite{gong2022ssast, baade2022mae, xu2022masked}. However, using L1 or L2 losses for reconstruction can result in a blurred image or a noisy audio signal. As He et al. suggest \cite{he2022masked}, improving the quality of MAE predictions can potentially lead to better representations for downstream tasks. 

    This paper introduces the vector quantized MAE for speech (VQ-MAE-S), a self-supervised model designed for emotion detection in speech signals. VQ-MAE-S is an adapted version of the Audio-MAE model proposed in \cite{gong2022ssast, baade2022mae, xu2022masked}. Unlike Audio-MAE, VQ-MAE-S operates on the discrete latent representation of a vector-quantized variational autoencoder (VQ-VAE) \cite{van2017neural} instead of the spectrogram representation. The pre-training of the model is performed on the VoxCeleb2 dataset \cite{chung2018voxceleb2}, and fine-tuning is carried out on several standard emotional speech datasets. We conduct several experiments to study the impact of different model design choices (e.g., masking ratio, masking strategy, patch size, etc.). The experimental results demonstrate that the proposed VQ-MAE-S model yields improvements in SER performance compared to an MAE using raw spectrogram data and other state-of-the-art methods. The code and qualitative reconstruction results are available at \url{https://samsad35.github.io/VQ-MAE-Speech/}.

\section{Vector Quantized Masked Autoencoder for Speech}
\label{sec:methodology}

\begin{figure*}[t]
    \centering
    \includegraphics[width=\textwidth]{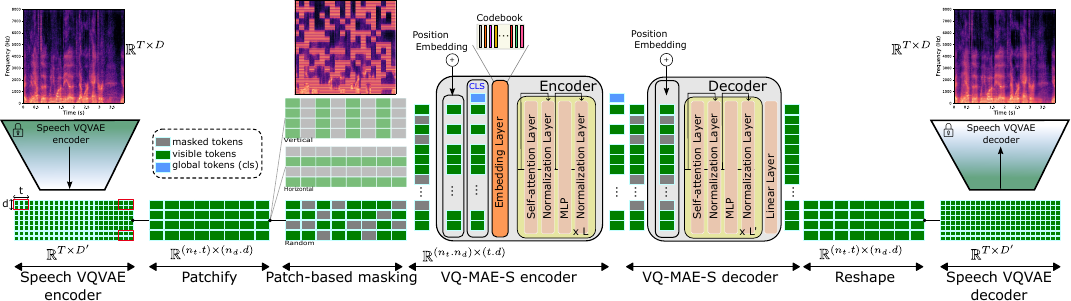}
    \caption{VQ-MAE-S model structure.}
    \label{fig:Overview}
\end{figure*}

This section presents the proposed self-supervised VQ-MAE-S model, which is represented in Figure~\ref{fig:Overview}. The model takes as input the power spectrogram of a speech signal, denoted by $\mathbf{x} \in \mathbb{R}_+^{T \times D}$, where $T$ and $D$ correspond to the time and frequency dimensions. 

\subsection{Vector quantized variational autoencoder}

The proposed self-supervised approach builds upon the discrete latent representation of a VQ-VAE \cite{van2017neural} assumed to be pre-trained and frozen (more details on the pre-training are given in Section~\ref{sec:pretrain_setting}). The VQ-VAE encoder is used to obtain a compressed and quantized representation $\mathbf{x}_q \in \mathbb{Z}^{T \times D'}$ of the input speech power spectrogram $\mathbf{x} \in \mathbb{R}_+^{T \times D}$. Each entry of $\mathbf{x}_q$ corresponds to the index of a vector in the VQ-VAE codebook. The quantized representation $\mathbf{x}_q$ keeps the aspect of a time-frequency representation, as the VQ-VAE is designed to be fully convolutional on the frequency axis and to process spectrogram frames independently. $\mathbf{x}_q$ can thus be seen as a discrete and compressed representation of the speech power spectrogram $\mathbf{x}$, where compression occurs along the frequency axis ($D' \ll D$). As illustrated in Figure~\ref{fig:Overview} and discussed in the next subsections, the proposed self-supervised learning approach operates on this discrete and compressed representation before reconstruction with the VQ-VAE decoder.

\subsection{Tokens creation, masking, and embedding}
\label{sec:mask}

\myparagraph{Discrete index tokens} As illustrated in Figure~\ref{fig:Overview}, we build \emph{discrete index tokens} from the quantized representation $\mathbf{x}_q$ by dividing it into non-overlapping patches of size ($t \times d$). This leads to $n_t = \frac{T}{t}$ and $n_d = \frac{D'}{d}$ patches horizontally and vertically, respectively. The representation $\mathbf{x}_q \in \mathbb{Z}^{(n_t \cdot t) \times (n_d \cdot d)}$ is then reshaped to $\mathbb{Z}^{(n_t \cdot n_d) \times (t \cdot d)}$ and used as input to the masking process. 

\myparagraph{Masking} We apply masking to the $(n_t \times n_d)$ time-frequency grid of discrete index tokens, each of which has a dimension of $t \cdot d$. The masked tokens are then replaced with a trainable vector. We explore three different types of patch-based masking, where tokens are masked randomly along the time-frequency, time, or frequency dimensions. Additionally, we test the frame masking strategy that involves masking the entire discrete frames of $\mathbf{x}_q$ (columns) instead of the patches. The effectiveness of these strategies is experimentally evaluated in the context of SER, including varying the ratio of masked tokens from $50$\% to $90$\%.

\myparagraph{Continuous embedding tokens} The discrete index tokens correspond to the indices obtained through the quantization step of the pretrained VQ-VAE encoder. Before being input to the VQ-MAE-S encoder, these discrete tokens are replaced with trainable continuous embedding vectors taken from a codebook of dimension $\mathbb{R}^{k \times e}$, where $k$ is the number of codes in the codebook, and $e$ is the dimension of each code. This is simply achieved by replacing the $t \cdot d$ indices of a discrete token by the corresponding $t \cdot d$ vectors of dimension $e$ in the codebook. The codebook is initialized by the pretrained VQ-VAE codebook and it can be held frozen or fine-tuned. After this embedding process (represented by the orange block in Figure\ref{fig:Overview}), the discrete index tokens in $\mathbb{Z}^{(n_t \cdot n_d) \times (t \cdot d)}$ are transformed into continuous embedding tokens in $\mathbb{R}^{(n_t \cdot n_d) \times (t \cdot d \cdot e)}$. In the experiments, we will investigate the effect of varying the embedding size $e$ on the SER performance.

\subsection{VQ-MAE-S encoder, decoder, and loss function}


\myparagraph{Encoder} The VQ-MAE-S encoder, similar to the ViT architecture \cite{dosovitskiy2020image}, consists of a single Transformer encoder \cite{vaswani2017attention}. This encoder is a stack of $L$ residual blocks that includes a self-attention layer, a normalization layer, and a Multi-layer Perceptron (MLP) block. Trained position embeddings are added for each token \cite{devlin2018bert}. As in \cite{he2022masked}, the encoder inputs are only the visible tokens; this is to learn a representation that relies on the context. In addition, we add a global trainable token $\texttt{[CLS]}$ as in \cite{devlin2018bert}, which will be used for SER tasks. 
Since most of the tokens are masked and only the visible tokens are fed to the encoder, this resolves the quadratic complexity issue inherent in transformer models with respect to the sequence length \cite{he2022masked}. As a result, the number of attention blocks in the encoder can be increased without experiencing computational inefficiencies.

\myparagraph{Decoder} The VQ-MAE-S decoder takes in both visible and masked tokens along with an additional position embedding. It consists of attention blocks similar to the encoder, but with fewer consecutive blocks ($L'$) compared to the encoder ($L > L'$). The decoder also includes a linear layer at the end that maps to the size of the VQ-VAE codebook. The output of this linear layer corresponds to the logits of the discrete index tokens. After applying a $\max$ operation, we obtain a reconstruction $\mathbf{x}_q^{rec}$ of the indices $\mathbf{x}_q$ that were provided by the VQ-VAE encoder.

\myparagraph{Loss function} To train the VQ-MAE-S model, we minimize the cross-entropy loss applied only to the masked discrete index tokens:
\begin{equation}
    \mathcal{L}_{\text{\tiny{VQ-MAE-S}}} = \texttt{cross-entropy}\big(\mathbf{x}_q(\Omega_\mathcal{M}),~ \mathbf{x}_q^{rec}(\Omega_\mathcal{M})\big),
\end{equation}
where $\mathbf{x}_q(\Omega_\mathcal{M})$ and $\mathbf{x}_q^{rec}(\Omega_\mathcal{M})$ represent the masked tokens in $\mathbf{x}_q$ and their reconstructions, respectively.

\begin{table*}[t]
    \centering
    \begin{minipage}{.45\linewidth}
      \centering
        \caption{Performance of VQ-MAE-S-\emph{12} on RAVDESS-Speech. Masking method: Random time-frequency masking (Patch-\emph{tf}), Ratio: 80\%, freeze refers to the freezing of the encoder of VQ-MAE-S.}
        \label{tab:finetuning}
            \begin{tabular}{c|ccc}
                Method          & Pre-train  & freeze  & Acc. \small{(\%)} \\ \hline
                VQ-MAE-S-\emph{12}        & \ding{55}               & \ding{55}             & 26.8  \\
                VQ-MAE-S-\emph{12}        & \checkmark             & \checkmark            & 57.6  \\
                VQ-MAE-S-\emph{12}        & \checkmark             & \ding{55}             & \textbf{76.7}  \\
                SpecMAE-\emph{12} & \checkmark            & \ding{55}             &   52.2            
            \end{tabular}
    \end{minipage} 
    \hfill
    \begin{minipage}{.5\linewidth}
      \caption{Performance of VQ-MAE-S-\emph{12} on RAVDESS-Speech for different masking strategies with a ratio of 80\%.}
      \label{tab:mask}
      \centering
        \begin{tabular}{c|cc}
                Masking method     & Acc. \small{(\%)} & f1-score \small{(\%)} \\ \hline
                Random frame masking (Frame) & \textbf{80.8} & \textbf{80.5}\\ 
                Random frequency masking (Patch-\emph{f})    & 65.7           &     65.0     \\
                Random time masking (Patch-\emph{t}) & 68.5           & 68.7         \\
                Random time-frequency masking  (Patch-\emph{tf}) & \textbf{76.7}           &     \textbf{75.9}         
            \end{tabular}
    \end{minipage}%
    \hfill
    \begin{minipage}{.45\linewidth}
      \centering
        \caption{Performance of VQ-MAE-S-\emph{12} on RAVDESS-Speech for different continuous embedding sizes. Masking method: Random time-frequency masking (Patch-\emph{tf}); Ratio: 80\%}
        \label{tab:dim}
            \begin{tabular}{c|ccc}
                Token dim.     & Parm. \small{(M)} & Acc. \small{(\%)} & f1-score \small{(\%)}\\ \hline
                $(t \cdot d \cdot (e=4))$=160                & $\approx$5 & 70.9           &  70.1        \\
                $(t \cdot d \cdot (e=8))$=320                & $\approx$20 &\textbf{76.7}           &     \textbf{75.9}       \\
                $(t \cdot d \cdot (e=16))$=640                & $\approx$80 & 75.4           &   75.3     
            \end{tabular}
    \end{minipage} 
    \hfill
    \begin{minipage}{.5\linewidth}
      \centering
        \caption{Performance of VQ-MAE-S on RAVDESS-Speech for different encoder depths. Masking method: Random time-frequency masking (Patch-\emph{tf}); Ratio: 80\%.}
        \label{tab:depth}
            \begin{tabular}{c|ccc}
                Encoder Depth     & Parm. \small{(M)} & Acc. \small{(\%)} & f1-score \small{(\%)}\\ \hline
                VQ-MAE-S-\emph{6}  & $\approx$12 & 65.8           &    65.3      \\
                VQ-MAE-S-\emph{12} & $\approx$20 & 76.7           &    75.9         \\
                VQ-MAE-S-\emph{16} & $\approx$25 & \textbf{79.2}           &  \textbf{79.4}      \\
                VQ-MAE-S-\emph{20} & $\approx$30 & 76.8 &  76.2      
            \end{tabular}
    \end{minipage}%
    
\end{table*}

\section{Experiments}
\subsection{Pre-training, fine-tuning, and evaluation}   

\subsubsection{Pre-training of VQ-MAE-Speech}
\label{sec:pretrain_setting}
To pre-train VQ-MAE-S, we used the VoxCeleb2 dataset \cite{chung2018voxceleb2}, which provides an extensive collection of audio speech data from open-source media, including speech segments corrupted by various real-world noises. We restricted our use of the dataset to a subset of around 1000 hours of audio-visual speech, including 2170 speakers.

The VQ-VAE model was trained on the same portion of the VoxCeleb2 dataset using short-time Fourier transform (STFT) power spectrograms ($\mathbf{x}$). The STFT is computed using a 64-ms Hann window (1024 samples) and a 70\% overlap, resulting in sequences of $D=513$-dimensional discrete Fourier coefficients. The VQ-VAE architecture is symmetrical with respect to the encoder and the decoder, with three 1D convolution (encoder) or transposed convolution (decoder) layers on the frequency axis and a residual convolution layer. The model processes each frame independently with no time dependency. For each speech power spectrogram frame of size $D=513$, the VQ-VAE compresses it into a discrete latent vector (a row of $\mathbf{x}_q$) of size $D'=64$. 
The VQ-VAE codebook contains $k=256$ codes of dimension $e=8$. Such a low dimension is chosen to increase the use of the different codes in the codebook, as in \cite{yuvector}. 

We evaluate the impact of different encoder architectures on the performance of VQ-MAE-S. The architecture is denoted as VQ-MAE-S-\emph{n}, where \emph{n} refers to the number of attention blocks in the encoder.  The decoder is fixed at four attention blocks. Each self-attention layer in these blocks is subdivided into four heads. The model is trained using the AdamW optimizer \cite{loshchilov2017decoupled} with a cosine scheduler to adjust the learning rate, with a 100-epoch warm-up period. The parameters of the optimizer, similar to \cite{he2022masked}, are $\beta_2=0.9$, $\beta_2=0.95$, and \texttt{weight\_decay}$=0.05$. The base learning rate follows the linear scaling rule \cite{goyal2017accurate} $\texttt{lr} = (\texttt{base\_lr} = 1e-3) \times (\texttt{batchsize} = 512) / 256$. We distributed the pre-training of VQ-MAE-S on 4 NVIDIA HGX A100. As in \cite{xu2022masked}, no data augmentation is performed. For more information on the architecture, please refer to our publicly available implementation.

\subsubsection{SER fine-tuning details}

Only the encoder of the VQ-MAE-S model is fine-tuned for SER. We propose two approaches: The first one uses the global token $\texttt{[CLS]}$ as input to a single linear layer, followed by a $\max$ operation to perform emotion classification.
The second approach, referred to as Query2Emo and inspired by \cite{liu2021query2label}, involves cross-attention between all sequence tokens as value/key and the emotion classes represented by trainable embedding as query. Query2Emo has a single attention block for both the encoder and decoder. For these two approaches, we use the AdamW optimizer \cite{loshchilov2017decoupled} with a cosine scheduler to adjust the learning rate and with a 40-epoch warm-up period. The parameters of the optimizer, similar to \cite{he2022masked}, are $\beta_2=0.9$, $\beta_2=0.95$, and \texttt{weight\_decay} $=0.05$. The base learning rate is \texttt{1e-4}. For the loss function, we adopt the asymmetric loss \cite{ridnik2021asymmetric} adapted for single labels instead of the cross entropy as it yields better results. 

\subsubsection{Emotional databases for fine-tuning and evaluation}

We fine-tune and evaluate the proposed approaches on four emotional speech audio databases.\\
\textbf{RAVDESS-Speech} \cite{livingstone2018ryerson}: This English database consists of 1440 audio files recorded by 24 professional actors and labeled with eight different emotional expressions (neutral, calm, happy, sad, angry, fearful, disgust, surprised). \\
\textbf{RAVDESS-Song} \cite{livingstone2018ryerson}: Same as the RAVDESS-Speech database, but utterances are sung \textit{a capella}. This database contains a total of 1012 audio files recorded by 23 actors and labeled with six emotions (neutral, calm, happy, sad, angry, and fearful). \\
\textbf{IEMOCAP} \cite{busso2008iemocap}: This database comprises approximately 12 hours of audio, annotated with several emotions, but only four emotions (neutral, happy, angry, and disgusted) have been retained to ensure a balanced distribution. It consists of dyadic sessions in which actors participate in improvisations or scripted scenarios.\\
\textbf{EMODB} \cite{burkhardt2005database}: The German EMODB database consists of 535 utterances spoken by ten professional speakers. It includes seven emotions (anger, boredom, anxiety, happiness, sadness, disgust, and neutral). 

For a fair comparison with the literature, we perform \emph{5}-fold cross-validation by separating the speakers' identity between the fine-tuning phase and the evaluation phase.

\begin{table*}[t]
\centering
\caption{Overall results (accuracy (\%) and f1-score (\%)) on the four evaluation databases.}
\label{tab:my-table}
\begin{tabular}{c|cc|cc|cc|cc}
\texttt{DATASET} & \multicolumn{2}{c|}{\texttt{RAVDESS-Speech}} & \multicolumn{2}{c|}{\texttt{RAVDESS-Song}} & \multicolumn{2}{c|}{\texttt{IEMOCAP}} & \multicolumn{2}{c}{\texttt{EMODB}} \\ \hline
Metrics & Accuracy           & f1-score          & Accuracy          & f1-score         & Accuracy           & f1-score           & Accuracy          & f1-score          \\ \hline
Self-attention audio \cite{chumachenko2022self}        & 58.3               &   -          & -              &     -       &      -         &      -        &        -      &          -     \\
SSAST \cite{gong2022ssast} (Patch-\emph{tf})        &  -      &   -          & -              &     -       &      54.3         &      -        &        -      &          -     \\
MAE-AST \cite{baade2022mae} (Patch-\emph{tf}) &  -      &   -          & -              &     -       &      58.6         &      -        &        -      &          - \\
SpecMAE-12 (Patch-\emph{tf})        &  52.2      &   52.0          & 54.5              &     53.9       &      46.7         &      45.9        &        57.2      & 57.0    \\ \hline
VQ-MAE-S-\emph{12}  (Patch-\emph{tf})      &   76.7             &      75.9       &        84.0     &    84.0        &    61.9           &   61.2           &    85.7          &      85.8       \\
VQ-MAE-S-\emph{12} (Patch-\emph{tf}) + Query2Emo        & 78.2               &   77.5          &       83.7       &      83.4      &    63.1           &   62.5           &      88.4        &  88.3           \\
VQ-MAE-S-\emph{12} (Frame)       & 80.8               &   80.5          &       84.2       &      84.3      &    65.2           &   64.9           &      87.0        &  87.0       \\  
VQ-MAE-S-\emph{12} (Frame) + Query2Emo      & \textbf{84.1}               &   \textbf{84.4}          &       \textbf{85.8}       &      \textbf{85.7}      &    \textbf{66.4}           &   \textbf{65.8}           &      \textbf{90.2}        &  \textbf{89.1}          
        
\end{tabular}
\end{table*}

\subsection{Performance on speech emotion recognition}
Table~\ref{tab:finetuning} highlights the importance of pre-training and fine-tuning the VQ-MAE-S-\emph{12} model for SER. Pre-training significantly improves the SER performance, with the accuracy increasing from $28.6$\% to $76.7$\% on the RAVDESS-Speech dataset. Fine-tuning the encoder is also crucial, as freezing the encoder results in a $19.1$\% accuracy loss. We also compared our approach with SpecMAE-\emph{12}, an MAE model that directly uses speech power spectrogram patches and aims to reconstruct masked patches using L2 loss. Our approach outperforms SpecMAE-\emph{12} by $24.5$\% in terms of accuracy, which clearly shows the benefit of working on the discrete representation of the VQ-VAE instead of the raw spectrogram representation for training the MAE.

Table~\ref{tab:my-table} compares the SER performance (accuracy and F1-score) of the proposed VQ-MAE-S model (with classification from the $\texttt{[CLS]}$ token), its improved version VQ-MAE-S + Query2Emo, the SpecMAE-12 baseline, and three state-of-the-art methods: SSAST \cite{gong2022ssast}, MAE-AST \cite{baade2022mae}, and a supervised self-attention-based approach \cite{chumachenko2022self} on the four evaluation databases. Two configurations of masking are considered: random frame masking (Frame) and random time-frequency patch masking (Patch-\emph{tf}). The results indicate that the proposed models outperform all other methods across all databases.
For random time-frequency patch masking (Patch-\emph{tf}), VQ-MAE-S achieves $15.2$\% better accuracy than SpecMAE, $7.6$\% better accuracy than SSAST, and $3.3$\% better accuracy than MAE-AST on the IEMOCAP dataset. The accuracy improvement over the supervised method on the RAVDESS-Speech database is of $18.4$\%. Query2Emo also contributes to the SER performance, with a gain of $1.5$\%, $1.2$\%, and $2.7$\% compared to VQ-MAE-S (Patch-\emph{tf}) alone on RAVDESS-Speech, IEMOCAP, and EMODB, respectively.

We conducted several experiments to assess the importance of several hyperparameters of the proposed VQ-MAE-S model, which are presented and discussed in the following paragraphs.

\myparagraph{Impact of the masking strategy} The choice of the masking strategy in MAE-based self-supervised models can have a significant impact on the performance of auxiliary tasks \cite{he2022masked}. As shown in Table~\ref{tab:mask}, which compares the performance of emotion recognition across four masking types discussed in Section~\ref{sec:mask}, random frame-based masking outperforms random patch-based masking. In particular, the frame-based approach achieves performance gains of $15.1$\%, $12.3$\%, and $4.1$\% compared to random patch-based masking on the time, frequency, and time-frequency axes, respectively. These results are consistent with those reported in prior studies \cite{gong2022ssast, baade2022mae}, and are highlighted in Table~\ref{tab:my-table} for the four evaluated databases.

\begin{figure}[h]
    \centering
    \includegraphics[width=0.45\textwidth]{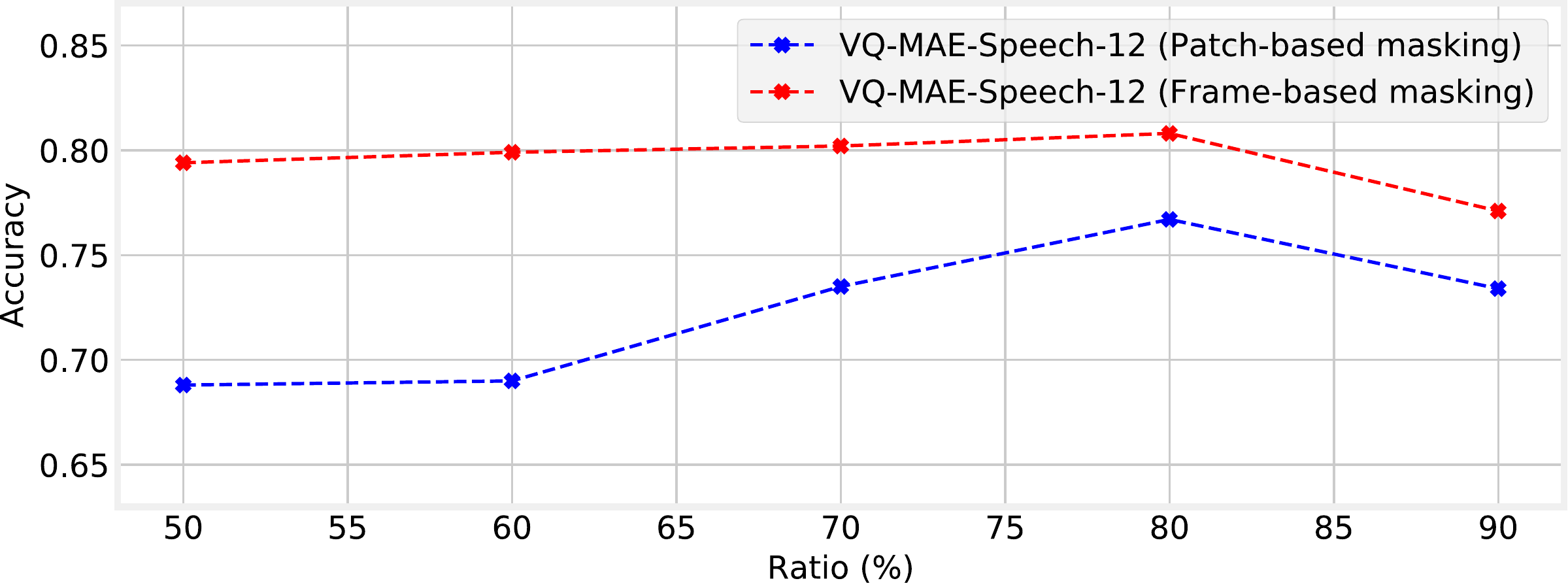}
    \caption{Impact of the masking ratio for VQ-MAE-S-\emph{12}.}
    \label{fig:ratio}
\end{figure}
\myparagraph{Impact of the masking ratio} Figure \ref{fig:ratio} shows the relationship between masking ratio (x-axis) and SER accuracy (y-axis) for the patch (blue) and frame (red) masking using VQ-MAE-S. The results indicate that patch masking achieves maximum accuracy at around $80\%$ masking ratio, after which performance declines. The trend can be attributed to the complexity of the task, where higher masking ratios improve representation learning and SER accuracy. In contrast, the performance of frame-based masking remains relatively stable. However, pushing the complexity too high (e.g., $90\%$) leads to a loss of relevant contextual information and decreased SER accuracy.

\myparagraph{Impact of the encoder depth}
The impact of the encoder depth on SER performance is presented in Table~\ref{tab:depth}, where the number of attention blocks is varied. The results show that, to some extent, increasing the number of blocks in the encoder leads to improved performance. When the number of blocks becomes very high (in this case, VQ-MAE-S-\emph{20}), there is no further improvement, and performance actually decreases ($-2.4$\% accuracy) compared to VQ-MAE-S-\emph{16}.

\myparagraph{Impact of the continuous embedding token size} The impact of the continuous embedding token size on the SER performance on the RAVDESS-Speech dataset is shown in Table~\ref{tab:dim}, by varying the dimension ($e$) of the codes in the dictionary. Although we did not conduct an extensive study on this parameter, we tested three exponentially increasing token sizes ($160, 320, 640$). It is observed that using a size of $320$ leads to an improvement in SER performance compared to sizes $160$ and $640$, with an increase of $+5.8$\% accuracy and $+1.3$\% accuracy, respectively.

\begin{figure}[h]
    \centering
    \includegraphics[width=0.45\textwidth]{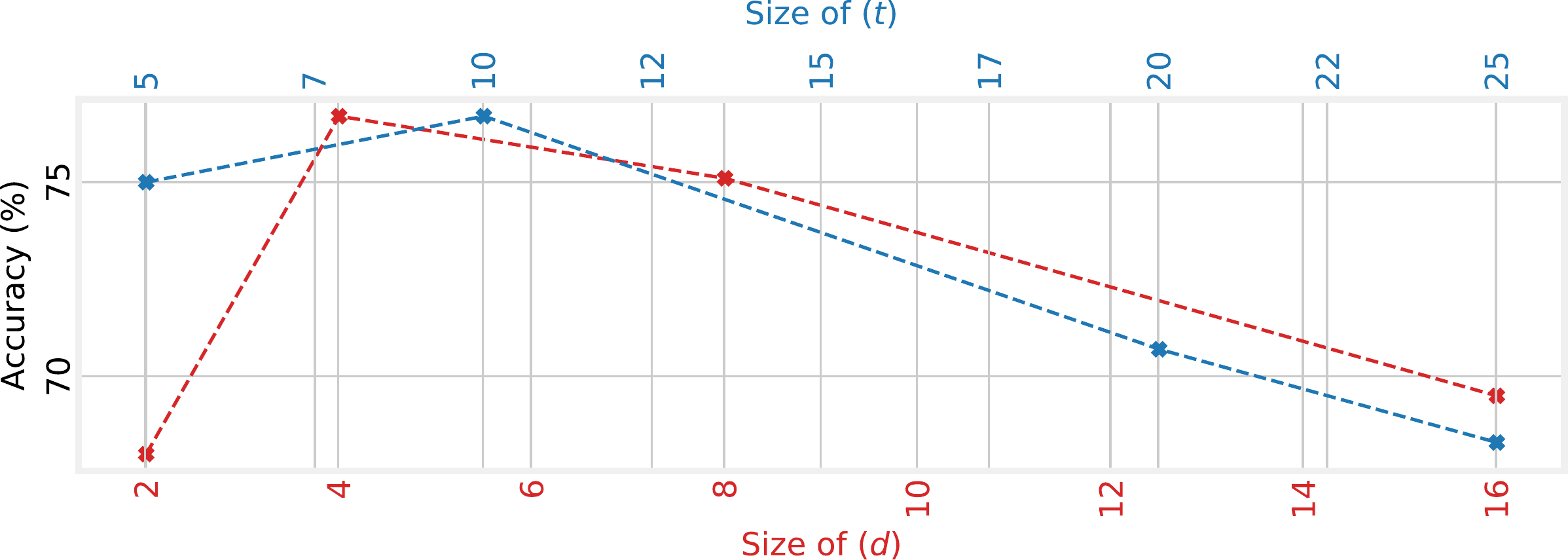}
    \caption{Impact of the discrete index token size ($t,~d$) on SER.}
    \label{fig:patch}
\end{figure}

\myparagraph{Impact of the discrete index token size}
Figure~\ref{fig:patch} illustrates the impact of the dimensions of the discrete index tokens ($t$ and $d$) on SER performance on the RAVDESS-Speech dataset. $d$ represents the discrete token size on the frequency axis, while $t$ represents the discrete token size on the time axis. Our study reveals that both $t$ and $d$ significantly affect SER performance. Therefore, it is important to carefully select the values of ($t,~d$). Based on our experiment, we suggest fixing them to ($t=10,~d=4$).


\section{Conclusion}
\label{sec:conclusion}
VQ-MAE-S is a novel approach that uses a pre-trained VQ-VAE model to adapt the MAE for speech representation learning, using discrete tokens obtained from the quantization step of the VQ-VAE. Our method outperforms other MAE methods based on spectrograms or mel-spectrograms for SER on several standard datasets. Several experiments also highlighted the impact of VQ-MAE-S hyperparameters on SER performance. For future work, we plan to investigate other masking strategies to further improve SER performance, and we aim to combine the MAE with contrastive methods to improve the learned audio representation and achieve even better SER performance.
\clearpage



\bibliographystyle{IEEEbib}
\bibliography{strings}

\begin{thebibliography}{10}

\bibitem{el2011survey}
Moataz El~Ayadi, Mohamed~S Kamel, and Fakhri Karray,
\newblock ``Survey on speech emotion recognition: Features, classification
  schemes, and databases,''
\newblock {\em Pattern recognition}, vol. 44, no. 3, pp. 572--587, 2011.

\bibitem{akccay2020speech}
Mehmet~Berkehan Ak{\c{c}}ay and Kaya O{\u{g}}uz,
\newblock ``Speech emotion recognition: Emotional models, databases, features,
  preprocessing methods, supporting modalities, and classifiers,''
\newblock {\em Speech Communication}, vol. 116, pp. 56--76, 2020.

\bibitem{wang2021fine}
Yingzhi Wang, Abdelmoumene Boumadane, and Abdelwahab Heba,
\newblock ``A fine-tuned wav2vec 2.0/hubert benchmark for speech emotion
  recognition, speaker verification and spoken language understanding,''
\newblock {\em arXiv preprint arXiv:2111.02735}, 2021.

\bibitem{chen2021exploring}
Li-Wei Chen and Alexander Rudnicky,
\newblock ``Exploring wav2vec 2.0 fine-tuning for improved speech emotion
  recognition,''
\newblock {\em arXiv preprint arXiv:2110.06309}, 2021.

\bibitem{gong2022ssast}
Yuan Gong, Cheng-I Lai, Yu-An Chung, and James Glass,
\newblock ``Ssast: Self-supervised audio spectrogram transformer,''
\newblock in {\em AAAI Conference on Artificial Intelligence}, 2022, vol.~36,
  pp. 10699--10709.

\bibitem{pepino2021emotion}
Leonardo Pepino, Pablo Riera, and Luciana Ferrer,
\newblock ``Emotion recognition from speech using wav2vec 2.0 embeddings,''
\newblock {\em International Speech Communication Association (Interspeech)},
  pp. 3400--3404, 2021.

\bibitem{liu2022audio}
Shuo Liu, Adria Mallol-Ragolta, Emilia Parada-Cabaleiro, Kun Qian, Xin Jing,
  Alexander Kathan, Bin Hu, and Bjoern~W Schuller,
\newblock ``Audio self-supervised learning: A survey,''
\newblock {\em in Patterns}, vol. 3, no. 12, pp. 100616, 2022.

\bibitem{zhang2022survey}
Chaoning Zhang, Chenshuang Zhang, Junha Song, John Seon~Keun Yi, Kang Zhang,
  and In~So Kweon,
\newblock ``A survey on masked autoencoder for self-supervised learning in
  vision and beyond,''
\newblock {\em arXiv preprint arXiv:2208.00173}, 2022.

\bibitem{macary2021use}
Manon Macary, Marie Tahon, Yannick Est{\`e}ve, and Anthony Rousseau,
\newblock ``On the use of self-supervised pre-trained acoustic and linguistic
  features for continuous speech emotion recognition,''
\newblock in {\em IEEE Spoken Language Technology Workshop (SLT)}. IEEE, 2021,
  pp. 373--380.

\bibitem{he2022masked}
Kaiming He, Xinlei Chen, Saining Xie, Yanghao Li, Piotr Doll{\'a}r, and Ross
  Girshick,
\newblock ``Masked autoencoders are scalable vision learners,''
\newblock in {\em IEEE/CVF Conference on Computer Vision and Pattern
  Recognition}, 2022, pp. 16000--16009.

\bibitem{devlin2018bert}
Jacob Devlin, Ming-Wei Chang, Kenton Lee, and Kristina Toutanova,
\newblock ``Bert: Pre-training of deep bidirectional transformers for language
  understanding,''
\newblock {\em Conference of the North {A}merican Chapter of the Association
  for Computational Linguistics: Human Language Technologies, Volume 1 (Long
  and Short Papers)}, pp. 4171--4186, 2019.

\bibitem{dosovitskiy2020image}
Alexey Dosovitskiy, Lucas Beyer, Alexander Kolesnikov, Dirk Weissenborn,
  Xiaohua Zhai, Thomas Unterthiner, Mostafa Dehghani, Matthias Minderer, Georg
  Heigold, Sylvain Gelly, et~al.,
\newblock ``An image is worth 16x16 words: Transformers for image recognition
  at scale,''
\newblock {\em in International Conference on Learning Representations (ICLR)},
  2020.

\bibitem{baade2022mae}
Alan Baade, Puyuan Peng, and David Harwath,
\newblock ``Mae-ast: Masked autoencoding audio spectrogram transformer,''
\newblock {\em arXiv preprint arXiv:2203.16691}, 2022.

\bibitem{xu2022masked}
Hu~Xu, Juncheng Li, Alexei Baevski, Michael Auli, Wojciech Galuba, Florian
  Metze, Christoph Feichtenhofer, et~al.,
\newblock ``Masked autoencoders that listen,''
\newblock {\em arXiv preprint arXiv:2207.06405}, 2022.

\bibitem{van2017neural}
Aaron Van Den~Oord, Oriol Vinyals, et~al.,
\newblock ``Neural discrete representation learning,''
\newblock {\em in Advances in neural information processing systems}, vol. 30,
  2017.

\bibitem{chung2018voxceleb2}
J~Chung, A~Nagrani, and A~Zisserman,
\newblock ``Voxceleb2: Deep speaker recognition,''
\newblock {\em in International Speech Communication Association
  (Interspeech)}, 2018.

\bibitem{vaswani2017attention}
Ashish Vaswani, Noam Shazeer, Niki Parmar, Jakob Uszkoreit, Llion Jones,
  Aidan~N Gomez, {\L}ukasz Kaiser, and Illia Polosukhin,
\newblock ``Attention is all you need,''
\newblock {\em in Advances in neural information processing systems}, vol. 30,
  2017.

\bibitem{yuvector}
Jiahui Yu, Xin Li, Jing~Yu Koh, Han Zhang, Ruoming Pang, James Qin, Alexander
  Ku, Yuanzhong Xu, Jason Baldridge, and Yonghui Wu,
\newblock ``Vector-quantized image modeling with improved vqgan,''
\newblock in {\em International Conference on Learning Representations}.

\bibitem{loshchilov2017decoupled}
Ilya Loshchilov and Frank Hutter,
\newblock ``Decoupled weight decay regularization,''
\newblock {\em in International Conference on Learning Representations (ICLR)},
  2017.

\bibitem{goyal2017accurate}
Priya Goyal, Piotr Doll{\'a}r, Ross Girshick, Pieter Noordhuis, Lukasz
  Wesolowski, Aapo Kyrola, Andrew Tulloch, Yangqing Jia, and Kaiming He,
\newblock ``Accurate, large minibatch sgd: Training imagenet in 1 hour,''
\newblock {\em arXiv preprint arXiv:1706.02677}, 2017.

\bibitem{liu2021query2label}
Shilong Liu, Lei Zhang, Xiao Yang, Hang Su, and Jun Zhu,
\newblock ``Query2label: A simple transformer way to multi-label
  classification,''
\newblock {\em arXiv preprint arXiv:2107.10834}, 2021.

\bibitem{ridnik2021asymmetric}
Tal Ridnik, Emanuel Ben-Baruch, Nadav Zamir, Asaf Noy, Itamar Friedman, Matan
  Protter, and Lihi Zelnik-Manor,
\newblock ``Asymmetric loss for multi-label classification,''
\newblock in {\em International Conference on Computer Vision (ICCV)}. IEEE
  Computer Society, 2021, pp. 82--91.

\bibitem{livingstone2018ryerson}
Steven~R Livingstone and Frank~A Russo,
\newblock ``The ryerson audio-visual database of emotional speech and song
  (ravdess): A dynamic, multimodal set of facial and vocal expressions in north
  american english,''
\newblock {\em in PloS one}, vol. 13, no. 5, pp. e0196391, 2018.

\bibitem{busso2008iemocap}
Carlos Busso, Murtaza Bulut, Chi-Chun Lee, Abe Kazemzadeh, Emily Mower, Samuel
  Kim, Jeannette~N Chang, Sungbok Lee, and Shrikanth~S Narayanan,
\newblock ``Iemocap: Interactive emotional dyadic motion capture database,''
\newblock {\em Language resources and evaluation}, vol. 42, pp. 335--359, 2008.

\bibitem{burkhardt2005database}
Felix Burkhardt, Astrid Paeschke, Miriam Rolfes, Walter~F Sendlmeier, Benjamin
  Weiss, et~al.,
\newblock ``A database of german emotional speech.,''
\newblock in {\em International Speech Communication Association
  (Interspeech)}, 2005, vol.~5, pp. 1517--1520.

\bibitem{chumachenko2022self}
Kateryna Chumachenko, Alexandros Iosifidis, and Moncef Gabbouj,
\newblock ``Self-attention fusion for audiovisual emotion recognition with
  incomplete data,''
\newblock in {\em International Conference on Pattern Recognition (ICPR)}.
  IEEE, 2022, pp. 2822--2828.

\end{thebibliography}

\end{document}